\documentclass[aps,preprintnumbers,eqsecnum,amsmath,amssymb,nofootinbib]{revtex4}
\usepackage[english]{babel}
\usepackage{url}
\usepackage{graphicx}
\usepackage{xcolor}
\usepackage{amsmath}
\usepackage{amssymb}
\usepackage{slashed}
\usepackage{multirow}

\begin{document}
\title{Analysis of $q_\mathrm{rec}^2$-distribution for $B\to K M_X$ and $B\to K^* M_X$ decays \\
in a scalar-mediator dark-matter scenario}
\author{Alexander Berezhnoy$^{a}$, Wolfgang Lucha$^{b}$ and Dmitri Melikhov$^{a,c,d}$}
\affiliation{
$^a$D.~V.~Skobeltsyn Institute of Nuclear Physics, M.~V.~Lomonosov Moscow State University, 119991, Moscow, Russia\\
$^b$Institute of High Energy Physics, Austrian Academy of Sciences, Nikolsdorfergasse 18, A-1050 Vienna, Austria\\
$^c$Joint Institute for Nuclear Research, 141980 Dubna, Russia\\
$^d$Faculty of Physics, University of Vienna, Boltzmanngasse 5, A-1090 Vienna, Austria}
\date{\today}
\begin{abstract}
We demonstrate that the scalar-mediator dark-matter scenario is
consistent with the experimental data on the decay $B\to K M_X$
and provides a good description of the shape of the observed excess.
Within this scenario, the interaction with dark-matter particles leads to approximately the same excess in
$\Gamma(B\to K^* M_X)$ and $\Gamma(B\to K M_X)$ compared to the Standard Model; also the differential distributions
of the excess events are similar in shape in the variable $q_\mathrm{rec}^2$ measured by experiment.
\end{abstract}
\maketitle

\section{Introduction}
 A recent Belle II observation \cite{Belle-II:2023esi} of $B\to K M_X$ --- for which also the notation $B\to K \slashed{E}$, with $\slashed{E}$ the missing energy, is used --- at a level much exceeding the Standard-Model (SM) prediction for $B\to K\nu\bar\nu$ \cite{Allwicher:2023xba},
\begin{eqnarray}
 \label{belle}
{\cal B}(B^+\to K^+ M_X)=(2.3\pm 0.7)\times 10^{-5}\simeq (5.4\pm 1.5)\,{\cal B}(B^+\to K^+\nu\bar\nu)_{\rm SM},
\end{eqnarray}
opened the window for immediate discussions of possible new-physics effects capable to explain this result (see, e.g.,~the recent publications
\cite{Allwicher:2023xba, Athron:2023hmz, Bause:2023mfe, Felkl:2023ayn, Abdughani:2023dlr,Dreiner:2023cms, He:2023bnk,datta,Calibbi:2025rpx,Altmannshofer:2025eor, Zhang:2024hkn,Bhattacharya:2024clv,Altmannshofer:2024kxb,Davoudiasl:2024cee,Kim:2024tsm,Marzocca:2024hua,Bolton:2024egx,He:2024iju,Hou:2024vyw,Gabrielli:2024wys,Loparco:2024olo,Ho:2024cwk,Fridell:2023ssf,McKeen:2023uzo,Altmannshofer:2023hkn,Wang:2023trd,He:2025jfc,fajfer,Berezhnoy:2023rxx}).
One of the most popular discussed options is the decay into Dark-Matter (DM) particles \cite{Abdughani:2023dlr} with multiple scenarios for the content of these particles and the possible mediators.

In our previous paper~\cite{Berezhnoy:2023rxx}, we noted that combining the current Belle II result for the decay $B\to K M_X$ with the hypothesis of a DM origin of the enhancement of the decay width $\Gamma(B\to K M_X)$, where the DM particles couple to SM particles via a scalar-mediator field, leads to rigorous constraints on $\Gamma(B\to K^* M_X)$ that are independent of further details of the DM model.

Extending the results of \cite{Berezhnoy:2023rxx}, this analysis provides a study of the differential distributions in the decays $B\to K M_X$ and $B\to K^* M_X$ in the experimental variable $q_\mathrm{rec}^2$ \cite{Belle-II:2023esi}. By confronting the calculated distributions with the data, we (i) obtain constraints on the numerical parameters of the DM model --- such as the mass and width of the scalar mediator, the mass of the light DM fermions, as well as the corresponding couplings --- and (ii) report that within the scalar mediator scenario, the $q_\mathrm{rec}^2$ differential distributions in $B\to K M_X$ and $B\to K^* M_X$ are equal to each other within a few-percent accuracy, independently of the numerical parameters of the DM model.

\section{$B\to (K,K^*) \bar\chi\chi$ decays via a scalar mediator $\phi$}
Following \cite{Berezhnoy:2023rxx}, we focus on a rather simple representative of the really vast class of DM models. To this end, we consider a rather popular model involving an interaction of DM fermions $\chi$ with the top quark $t$ by exchange of a scalar-mediator field $\phi$, governed by the interaction Lagrangian \cite{Batell:2009jf, Schmidt-Hoberg:2013hba}
\begin{eqnarray}
 \mathcal{L_{\rm int}} = - \frac{y m_t}{v} \phi\, \bar t t - \kappa \phi \bar \chi \chi.
  \label{eq:lagrangian}
\end{eqnarray}
The emerging effective Lagrangian encoding the flavor-changing neutral-current (FCNC) vertex $b\to s\phi$ then reads
\cite{Batell:2009jf, Schmidt-Hoberg:2013hba}
\begin{eqnarray}
\label{Leff}
\mathcal{L}_{b \rightarrow s\phi} = g_{b\to s\phi}\,\phi\, \bar s_L b_R + {\rm h.c.}, \qquad
 g_{b\to s\phi}=\frac{y m_b}{v} \frac{3 \sqrt{2} G_F m_t^2 V^*_{ts} V_{tb}}{16 \pi^2}.
\end{eqnarray}
This Lagrangian enters in the amplitudes controlling the $\phi$-mediated decay $B\to (K,K^*)\bar\chi\chi$:
\begin{eqnarray}
\label{Amp}
 A(B(p)\to K^{(*)}(p-q)\bar\chi(k)\chi(q-k))&=&-i\langle K^{(*)}(p-q)\bar\chi(k)\chi(q-k)|\mathcal{L}_{b\to s\phi}|B(p)\rangle \nonumber\\  
 & = &
 \langle \bar\chi\chi|\bar\chi\chi|0\rangle
 \kappa\frac{1}{M_\phi^2-q^2-i M_\phi \Gamma_\phi(q^2)}g_{b\to s\phi}\langle K^{(*)}(p-q)|\bar s_L b_R|B(p)\rangle,
\end{eqnarray}
with the abbreviation
\begin{equation}
\langle \bar\chi\chi|\bar\chi\chi|0\rangle\equiv \langle \bar\chi(k)\chi(q-k)|\bar\chi(0)\chi(0)|0\rangle.
\end{equation}
For the present analysis, we take the liberty to assume that $M_\phi> 2 m_\chi$ and furthermore that the mediator $\phi$ decays predominantly into the $\bar\chi\chi$ pair. As the consequence, we obtain a $q^2$-dependent width of the mediator $\phi$ which is calculated from the imaginary part of the fermion loop diagram with scalar vertices in the form (cf. \cite{gs,nachtmann})
\begin{equation}
\label{Gamma}
\Gamma_\phi(q^2)= \left(\frac{q^2-4m_\chi^2}{M_\phi^2-4m_{\chi}^2}\right)^{\frac{3}{2}}\frac{M_\phi}{\sqrt{q^2}}\; \Theta(q^2-4m_\chi^2)\; \Gamma_\phi^0, \qquad \Gamma_\phi^0=\frac{\kappa^2}{8\pi}M_\phi\left(1-\frac{4m_\chi^2}{M_\phi^2}\right)^{\frac{3}{2}}.
\end{equation}
Here, $q$ is the momentum of the outgoing $\chi\bar\chi$ pair of unobserved DM particles; $M_X^2\equiv q^2$ is the missing mass squared.

Equation~(\ref{Amp}) provides a simplified parametrization of the full propagator of the scalar particle, taking into account the resummation of the $\bar\chi\chi$ loops and neglecting the real parts of the loop diagrams.\footnote{Taking into account that the correction to the vector meson propagator due to the pion loops has precisely the same analytic expression as the correction to the scalar particle propagator due to the spin-1/2 fermion loops, one can directly use the real part of the loop diagram given by Eq.~(11) of \cite{gs}. However, this has a negligible impact on our results so we make use of a simplified expression (\ref{Amp}).}

Using the QCD equations of motion, the required amplitudes can be straightforwardly calculated, yielding
\begin{eqnarray}
\label{FF}
 \langle K|\bar s_L b_R|B \rangle &=&\frac{1}{2}\langle K|\bar s (1-\gamma_5) b|B\rangle=
 \frac{1}{2}\langle K|\bar s b|B\rangle=
 \frac{1}{2} \frac{M_B^2-M_K^2}{m_b-m_s}f_0^{B\to K}(q^2),\nonumber\\
 \langle K^*|\bar s_L b_R|B \rangle &=& \frac{1}{2}\langle K^*|\bar s(1-\gamma_5) b|B \rangle=
 -\frac{1}{2}\langle K^*|\bar s \gamma_5 b|B \rangle
 =-i (\epsilon q)\frac{M_{K^*}}{m_b+m_s}A_0^{B\to K^*}(q^2),
\end{eqnarray}
with well-known dimensionless form factors $f_0$ and $A_0$ parametrizing the amplitudes $\langle K|\bar s \gamma_\mu b|B\rangle$ and $\langle K^*|\bar s \gamma_\mu\gamma_5 b|B\rangle$ \cite{wsb}.

Using these amplitudes and the recursive formula for the phase space
\begin{eqnarray}
 d \Phi_3(M_B^2, p_K, p_\chi, p_{\bar \chi})&=&
 \Phi_2(M_B^2,p_K,p_\chi + p_{\bar \chi})\frac{d q^2}{2\pi} d\Phi_2(q^2,p_\chi, p_{\bar \chi})
 = \frac{\lambda^{1/2}(M_B^2,M_K^2,q^2)}{8 \pi M_B^2}\frac{d q^2}{16\pi^2} \sqrt{1-\frac{4 m_\chi^2}{q^2}},
\end{eqnarray}  
where $\lambda$ is defined by $\lambda(a,b,c)\equiv(a-b-c)^2 -4bc$, we can factorize the process into the decay of the $B$ meson to the $K$ meson and a virtual $\phi^*$, followed by the decay of $\phi^*$ to a $\chi\bar\chi$ pair. The $d\Gamma(B\to (K,K^*)\bar\chi\chi)/dq^2$ distributions can then be expressed as
\begin{eqnarray}
 \frac{d\Gamma}{dq^2}(B\to K \bar\chi\chi)&=&
 \frac{\sum_{\chi\,{\rm polar}}|A|^2}{2M_B}\frac{\lambda^{1/2}(M_B^2,M_K^2,q^2)}{8 \pi M_B^2} \frac{1}{16\pi^2} \sqrt{1-\frac{4 m_\chi^2}{q^2}}
 \nonumber\\  
&=&\frac{\lambda^{1/2}(M_B^2,M_K^2,q^2)}{16 \pi M_B^3}|\langle K|\bar s_L b_R|B \rangle|^2
 \frac{g_{b\to s\phi}^2\,\kappa^2}{(M_\phi^2-q^2)^2+M_\phi^2\Gamma_\phi^2(q^2)} \frac{1}{16\pi^2} \sqrt{1-\frac{4 m_\chi^2}{q^2}}\nonumber\\
 &&\times \sum_{\chi\,{\rm polar}}|\langle \bar\chi(k)\chi(q-k)|\bar\chi(0)\chi(0)|0\rangle|^2.
 \label{2.8}  
\end{eqnarray}
The expression for $B\to K^*\bar\chi\chi$ is obtained by the obvious replacements $K\to K^{*}$ and $M_K\to M_{K^*}$ in Eq.~(\ref{2.8}).

Taking into account that
\begin{equation}
\sum_{\chi\,{\rm polar}}|\langle \bar\chi(k)\chi(q-k)|\bar\chi(0)\chi(0)|0\rangle|^2=q^2-4m_\chi^2
\end{equation}  
and
\begin{equation}
\sum_{K^*\,{\rm polar}}|\langle K^*|\bar s_L b_R|B \rangle|^2= \frac{\lambda(M_B^2,M_{K^*}^2,q^2)}{4(m_b+m_s)^2} |A_0^{B\to K^*}(q^2)|^2,
\end{equation}
we obtain the following expressions for $d\Gamma(B\to K\bar\chi\chi)/dq^2$ and $d\Gamma(B\to K^*\bar\chi\chi)/dq^2$:
\begin{eqnarray}
\frac{d\Gamma}{dq^2}(B\to K \bar\chi\chi)&=&
\frac{\lambda^{1/2}(M_B^2,M_K^2,q^2)}{16 \pi M_B^3}
\frac{(M_B^2-M_K^2)^2 |f_0^{B\to K}(q^2)|^2}{4(m_b-m_s)^2}
\frac{g_{b\to s\phi}^2\,\kappa^2}{(M_\phi^2-q^2)^2+M_\phi^2\Gamma_\phi^2(q^2)}
 \frac{q^2}{16\pi^2} \left (1-\frac{4 m_\chi^2}{q^2}\right)^{3/2},\nonumber
\label{eq:dGdq2}\\
\frac{d\Gamma}{dq^2}(B\to K^* \bar\chi\chi)&=&\frac{\lambda^{3/2}(M_B^2,M_{K^*}^2,q^2)}{16 \pi M_B^3} \frac{|A_0^{B\to K^*}(q^2)|^2}{4(m_b+m_s)^2}
 \frac{g_{b\to s\phi}^2\,\kappa^2}{(M_\phi^2-q^2)^2+M_\phi^2\Gamma_\phi^2(q^2)}
 \frac{q^2}{16\pi^2} \left(1-\frac{4 m_\chi^2}{q^2}\right)^{3/2}.
\end{eqnarray}
A powerful probe of the DM scenario considered is provided by the ratio \cite{Berezhnoy:2023rxx}
\begin{eqnarray}
\label{R}
R^{(\phi)}_{K^*/K}(q^2)&=&\frac{d\Gamma(B\to K^*\chi\bar\chi)/dq^2}{d\Gamma(B\to K\chi\bar\chi)/dq^2}=
\frac{\lambda^{3/2}(M_B^2,M_{K^*}^2,q^2)}{\lambda^{1/2}(M_B^2,M_K^2,q^2)}
  \frac{|A_0^{B\to K^*}(q^2)|^2}{|f_0^{B\to K}(q^2)|^2}\frac{(m_b-m_s)^2}{(M_B^2-M_K^2)^2 (m_b+m_s)^2}.
\end{eqnarray}
Clearly, the dependence on the specific parameters of the DM model cancels out in this ratio. We shall see that this ratio remains largely unaffected by the averaging procedure adopted in the experiment.

The form factors $f_0$ and $A_0$ required by Eq.~(\ref{FF}) may be specified by implementing the results of \cite{ballP,ballV} in the~form of rather convenient parametrizations \cite{ms2000}:
\begin{eqnarray}
\label{f0}
f_0(q^2) &=& \frac{0.33}{1-0.7\, r_V+0.27\, r_V^2},\quad r_V\equiv q^2/M_{B_s^*}^2; \\
\label{A0}
A_0(q^2) &=& \frac{0.37}{(1-0.46\, r_P)(1-r_P)},\quad r_P\equiv q^2/M_{B_s}^2.
\end{eqnarray}
Numerically, these parametrizations yield, for the ratio of differential distributions in the reactions $B\to K \phi\to K\bar\chi\chi$ and $B\to K^* \phi\to K^*\bar\chi\chi$, the prediction depicted in Fig.~1 of \cite{Berezhnoy:2023rxx}. More recent parametrizations of these form factors (for instance, those from \cite{damir2023epjc,hpqcd2023}) entail, for $R^{(\phi)}_{K^*/K}$, an uncertainty of about 2\% \cite{Berezhnoy:2023rxx}.

 \section{Kinematics}
The Belle II experiment, as described in \citep{Belle-II:2023esi}, analyzes partially reconstructed events to increase the statistical sample. For such events, the direction of the $B$ meson cannot be determined. Therefore, instead of $q^2=(p_B-p_K)^2$ the variable $q_{\rm rec}^2$ is used:
 \begin{equation}
q^2_{\rm rec}= E_B^2+M_K^2-2E_B E_K,
 \end{equation}
where $E_B$ and $E_K$ are the energies of the $B$ and $K$ mesons in the center-of-mass frame of the $B\bar B$-meson pair produced in $\Upsilon(4S)$ decays.

The variable $q^2_{\rm rec}$ can be expressed through the three-momenta $\mathbf{p}_B$ and $\mathbf{p}_K$ of the $B$ and $K$ mesons, and then through the velocity $\mathrm{v}$ of the $B$ meson in the $B\bar B$ center-of-mass frame:
 \begin{equation}
q^2_{\rm rec}= E_B^2+M_K^2-2E_B E_K=q^2 + (E_B^2-M_B^2) - 2 (\mathbf{p}_K \cdot \mathbf{p}_B)=q^2+\frac{M_B^2 \mathrm{v}^2}{1-\mathrm{v}^2} -2 p_K^z \frac{M_B \mathrm{v}}{\sqrt{1-\mathrm{v}^2}},
\label{eq:q2rec}
 \end{equation}
where $\mathbf{p}_K^z$ is the projection of $\mathbf{p}_K$ onto the direction of the $B$-meson motion.

It is convenient to express $\mathbf{p}_K^z$ as follows:
\begin{equation}
\mathbf{p}_K^z=\frac{\cos \Theta \; \mathbf{p}_K^0-\mathrm{v}\;E_K^0}{\sqrt{1-\mathrm{v}^2}}.
\label{eq:pKz}
\end{equation}
Here, $\Theta$ is the angle between the direction of the $K$-meson motion in the $B$-meson rest frame and the direction of the $B$-meson motion in the $B\bar B$ center-of-mass frame, while $\mathbf{p}_K^0$ and $E_K^0$ are, respectively, the projection of the $K$-meson three-momentum onto the $B$-meson direction, and the $K$-meson energy in the $B$-meson rest frame:
\begin{equation}
\mathbf{p}_K^0=\frac{\lambda^{1/2}(M_B^2,M_K^2,q^2)}{2M_B}, \qquad E_K^0= \frac{M_B^2+M_K^2-q^2}{2M_B}.
\label{eq:pKz0}
\end{equation}

By combining Eqs.~(\ref{eq:q2rec}), (\ref{eq:pKz}) and (\ref{eq:pKz0}), $q_{\rm rec}^2$ becomes, in terms of $q^2$, $\Theta$ and $v$,
\begin{equation} q_{\rm rec}^2= \frac{q^2-M_K^2 \mathrm{v}^2- \cos \Theta \;\mathrm{v} \; \lambda^{1/2}(M_B^2,M_K^2,q^2)}
%
{1-\mathrm{v}^2}.
  \label{eq:qrec2_q2}
\end{equation}
This equation has an exact solution for $q^2$:
 \begin{multline}
 q^2=\frac{1}{1-\cos^2\Theta \; \mathrm{v}^2}
 \Bigg\{(1-\mathrm{v}^2)q_{\rm rec}^2+(1-\cos^2 \Theta) \mathrm{v}^2 M_K^2 -\cos^2\Theta\; \mathrm{v}^2 M_B^2 +\\+
 \cos\Theta\, \mathrm{v} \, \Big[ \lambda\Big(M_B^2, (1-\mathrm{v}^2)M_K^2,(1-\mathrm{v}^2)q_{\rm rec}^2\Big)-4 (1-\cos^2 \Theta)\mathrm{v}^2 M_B^2 M_K^2 \Big]^{1/2}\Bigg\}.
%
  \label{eq:q2_qrec2}
  \end{multline}
Neglecting terms proportional to $\mathrm{v}^2$, we obtain
 \begin{equation}
 q^2\approx
 q_{\rm rec}^2+ \cos\Theta\, \mathrm{v} \, \lambda^{1/2}(M_B^2,M_K^2,q_{\rm rec}^2).
\end{equation}
Further, upon neglecting $M_K$ we arrive at an even simpler relation between $q^2$ and $q_{\rm rec}^2$, which matches the previous expression within a few percent:
\begin{equation}
q^2\approx q_{\rm rec}^2+ (M_B^2-q_{\rm rec}^2) \mathrm{v} \cos \Theta.
\label{eq:q2_qrec2_approx}
\end{equation}
Since the $B$ meson decays uniformly over $\Theta$, we have
\begin{eqnarray}
\Gamma=\iint \frac{d^2 \Gamma (q^2,\cos \Theta)}{d q^2 d \cos \Theta} dq^2 d\cos \Theta= \frac{1}{2} \iint
 \frac{d \Gamma (q^2(q_{\rm rec}^2,\cos \Theta ))}{d q^2} \frac{dq^2(q_{\rm rec}^2,\cos \Theta )}{dq_{\rm rec}^2} d q_{\rm rec}^2 d\cos \Theta.
\end{eqnarray}
Thus,
 \begin{equation}
\frac{d\Gamma}{dq_{\rm rec}^2}= \frac{1}{2} \int
 \frac{d \Gamma (q^2(q_{\rm rec}^2,\cos \Theta ))}{d q^2} \frac{dq^2(q_{\rm rec}^2,\cos \Theta )}{dq_{\rm rec}^2} d\cos \Theta.
\end{equation}
Therefore, any decay characteristic ${\cal A}(q^2_{\rm rec})$ at a given $q^2_{\rm rec}$ is the average of ${\cal A}(q^2)$ over the range
 \begin{eqnarray}
  q^2 \in [q^2_{\rm rec}- \mathrm{v} \lambda^{1/2}(M_B^2,M_K^2,q^2_{\rm rec}); q^2+\mathrm{v} \lambda^{1/2}(M_B^2,M_K^2,q^2_{\rm rec})].
 \end{eqnarray}
In particular, the averaged differential decay rate has the form
 \begin{equation}
\frac{\overline{d\Gamma(q_{\rm rec}^2)}}{dq_{\rm rec}^2}\approx \Bigg\langle
 \frac{d \Gamma (q^2)}{d q^2}\Bigg \rangle_{q^2_{\rm rec}\pm \mathrm{v} \lambda^{1/2}(M_B^2,M_K^2,q^2_{\rm rec})}.
\label{eq:Grec2_Gq2}
\end{equation}
In order to give an idea of the size of the averaging interval, we note that $\mathrm{v} \lambda^{1/2}(M_B^2,M_K^2,0 \mbox{ GeV}^2)\approx 1.7$~GeV$^2$ and $\mathrm{v} \lambda^{1/2}(M_B^2,M_K^2,10\mbox{ GeV}^2)\approx 1$~GeV$^2$.

It is instructive to consider the real $\phi$-mediator with a negligible decay width, in which case one finds
\begin{equation}
\frac{d\Gamma}{dq^2}(B\to K^{(*)} \phi)=\Gamma(B\to K^{(*)} \phi) \cdot \delta(q^2-M_\phi^2).
\end{equation}
So, in this case, the distributions over $q_\mathrm{rec}^2$ become rectangular functions with a width of $2\mathrm{v} \lambda^{1/2}(M_B^2,M_{K^{(*)}}^2,q^2_{\rm rec})$:
\begin{equation}
\frac{d\Gamma}{dq_\mathrm{rec}^2}(B\to K^{(*)}\phi)
=\Gamma(B\to K^{(*)} \phi) \; \frac{\Theta(v \lambda^{1/2}(M_B^2,M_{K^{(*)}}^2,q^2_{\rm rec})-|q_\mathrm{rec}^2-M_\phi^2|)}{2\mathrm{v} \lambda^{1/2}(M_B^2,M_{K^{(*)}}^2,q^2_{\rm rec})}.
\end{equation}

\section{Data fitting}
The Belle II experiment does not apply efficiency corrections to its data~\cite{Belle-II:2023esi}. Consequently, in order to fit the~data, we adopt the $q^2$-dependent efficiency $\varepsilon(q^2)$ estimated in \cite{Fridell:2023ssf} (and provided by Fig.~2 of \cite{Fridell:2023ssf}),
\begin{equation}
\frac{\overline{d\Gamma_\mathrm{eff}(q_{\rm rec}^2)}}{dq_{\rm rec}^2}= \Bigg\langle
\varepsilon(q^2)\, \frac{d \Gamma (q^2)}{d q^2}\Bigg \rangle_{q^2_{\rm rec}\pm \lambda^{1/2}(M_B^2,M_K^2,q^2_{\rm rec})}.
\end{equation}
To determine the optimal fit parameters, we minimize the $\chi^2$ value, defined as
\begin{equation}
\label{chi2}
\chi^2=\sum_i\frac{(n^i_\mathrm{exp}-n^i_\mathrm{theor})^2}{(\Delta^i_{\rm exp})^2}+
\frac{(N_\mathrm{exp}-N_\mathrm{theor})^2}{\Delta_\mathrm{exp}^2},
\end{equation}
where the sum runs over the experimental bins, $n^i_\mathrm{exp}$ and $n^i_\mathrm{theor}$ are the measured and the predicted numbers of excess events over the SM expectation in the $i$-th bin, and $\Delta^{i}_{\rm exp}$ is the experimental error in the $i$-th bin; $N_{\rm exp}$ and $N_{\rm theor}$ are the full numbers of excess events,
$N_{\rm exp}=\sum_i n^i_\mathrm{exp}$ and $N_{\rm theor}=\sum_i n^i_\mathrm{theor}$,
and $\Delta_{\rm exp}^2=\sum_i (\Delta^{i}_{\rm exp})^2.$

The DM parameters to be determined by our analysis are thus $M_\phi$, $\Gamma_\phi^{0}$ and $m_\chi$, which determine the shape of the $q^2$-distribution, and the product of the couplings $g_{b\to s\phi}\kappa$ which determines the total yield of
the (invisible) $\bar\chi\chi$ pairs.
\section{Results}
\begin{figure}[t]
\begin{tabular}{ccc}
\includegraphics[width=8cm]{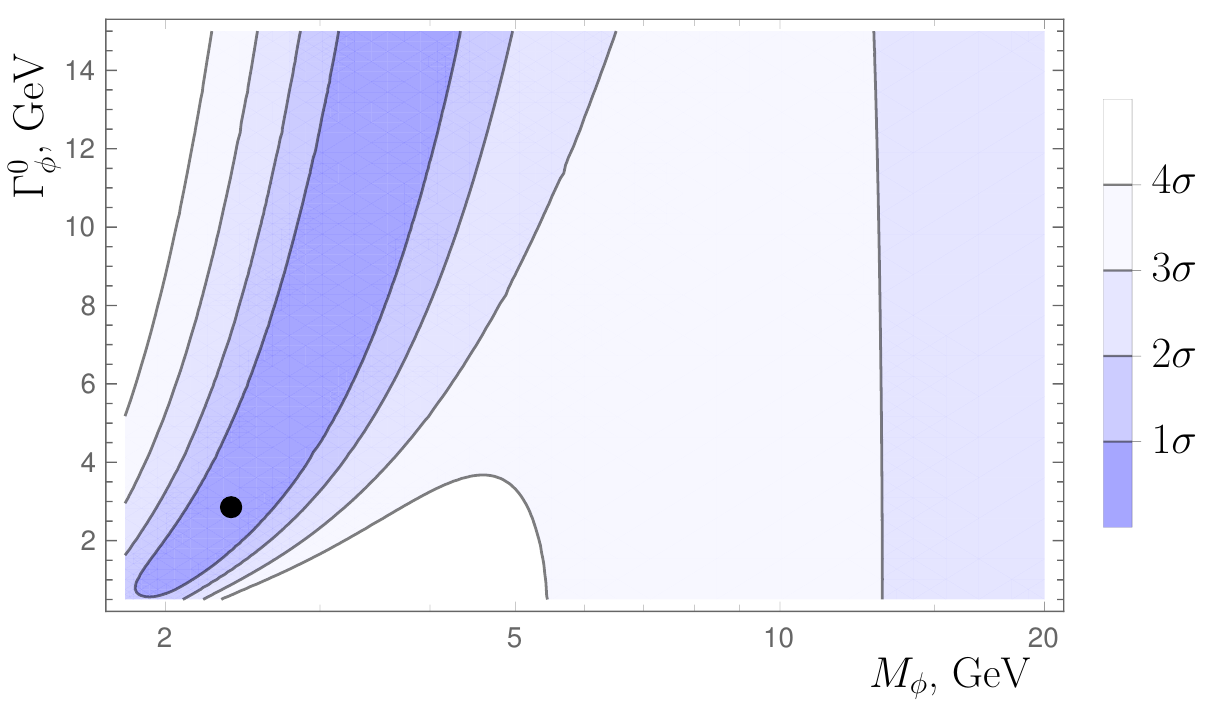} & ${}$\hspace{.5cm} ${}$ & \includegraphics[width=8cm]{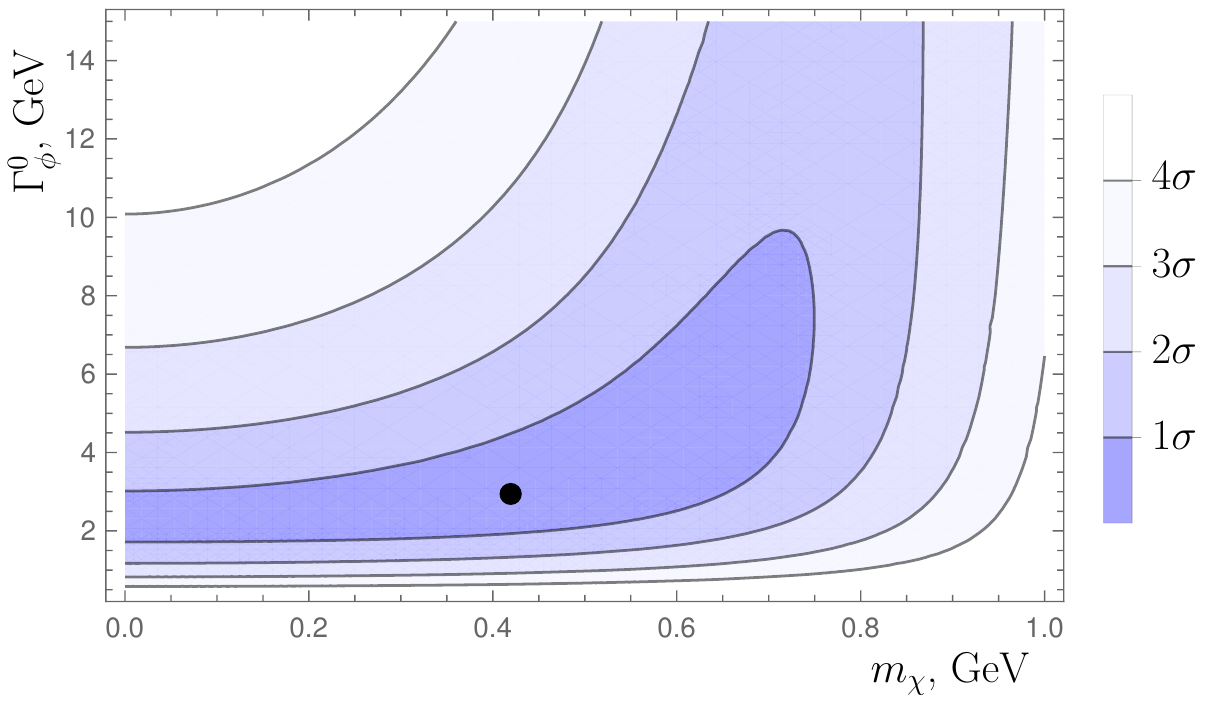}\\
(a) & & (b) \vspace{.5cm}\\
\includegraphics[width=8cm]{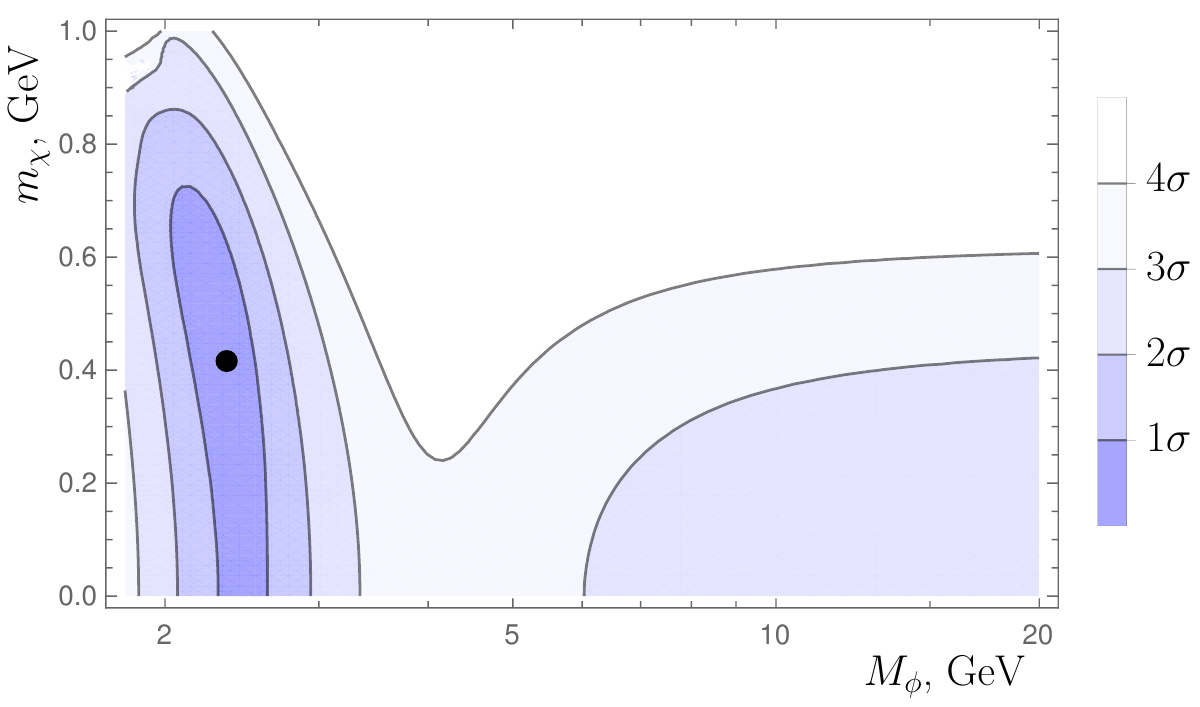} & &
\includegraphics[width=7.2cm]{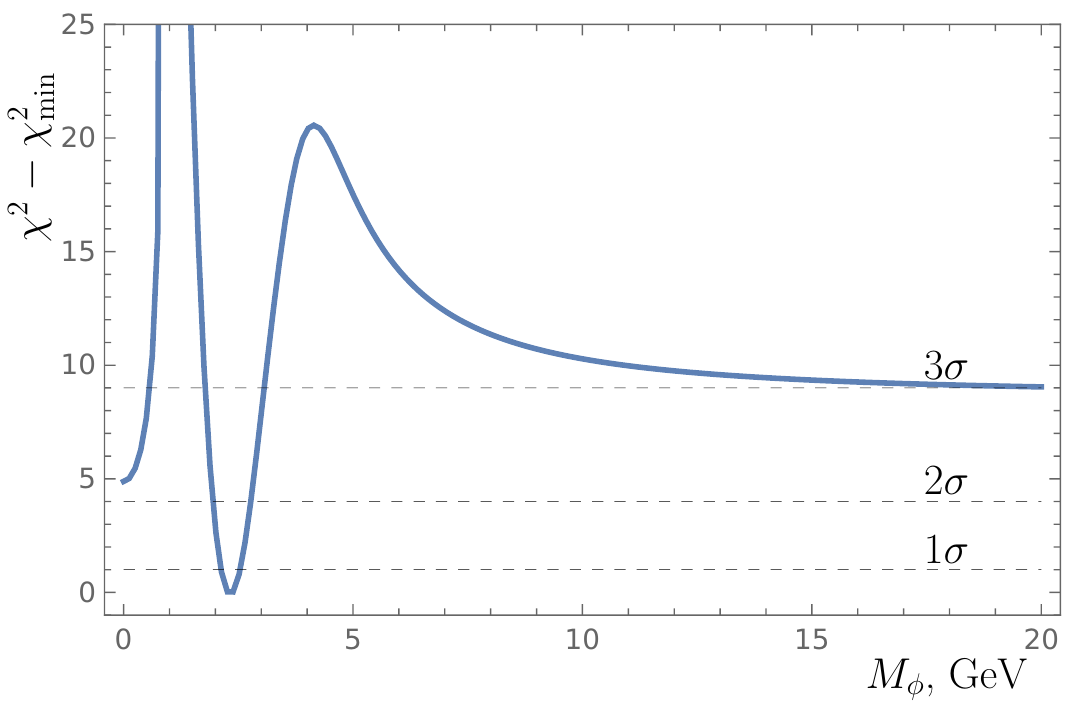}\\
(c) & & (d)\\
\includegraphics[width=7.2cm]{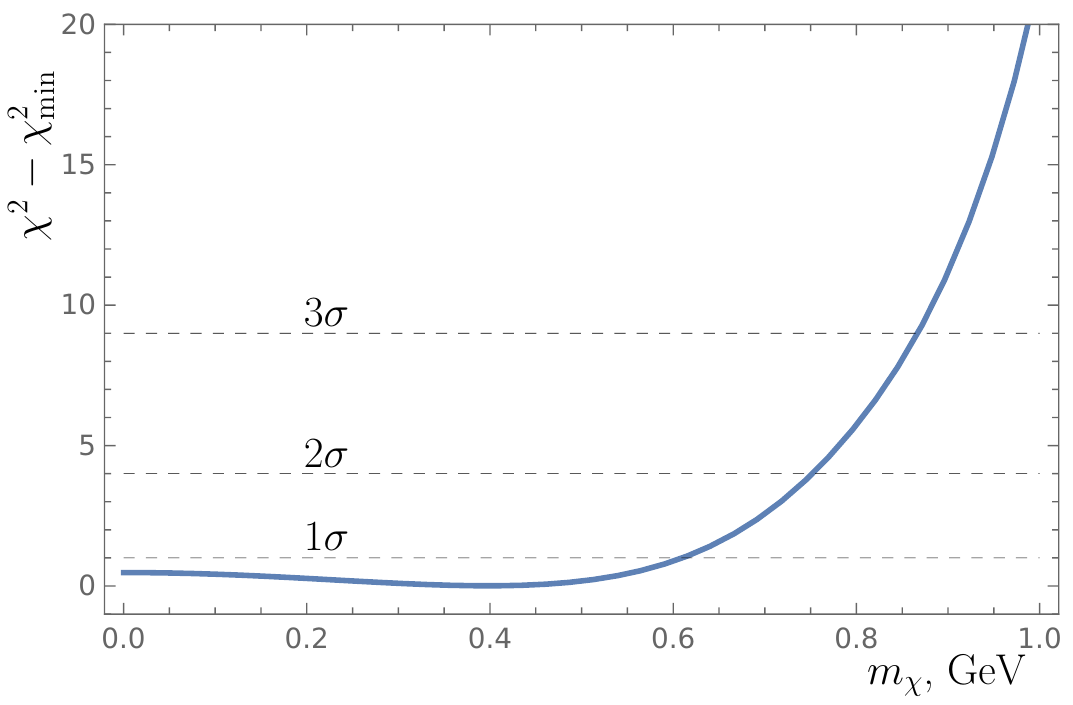} & & \includegraphics[width=7.2cm]{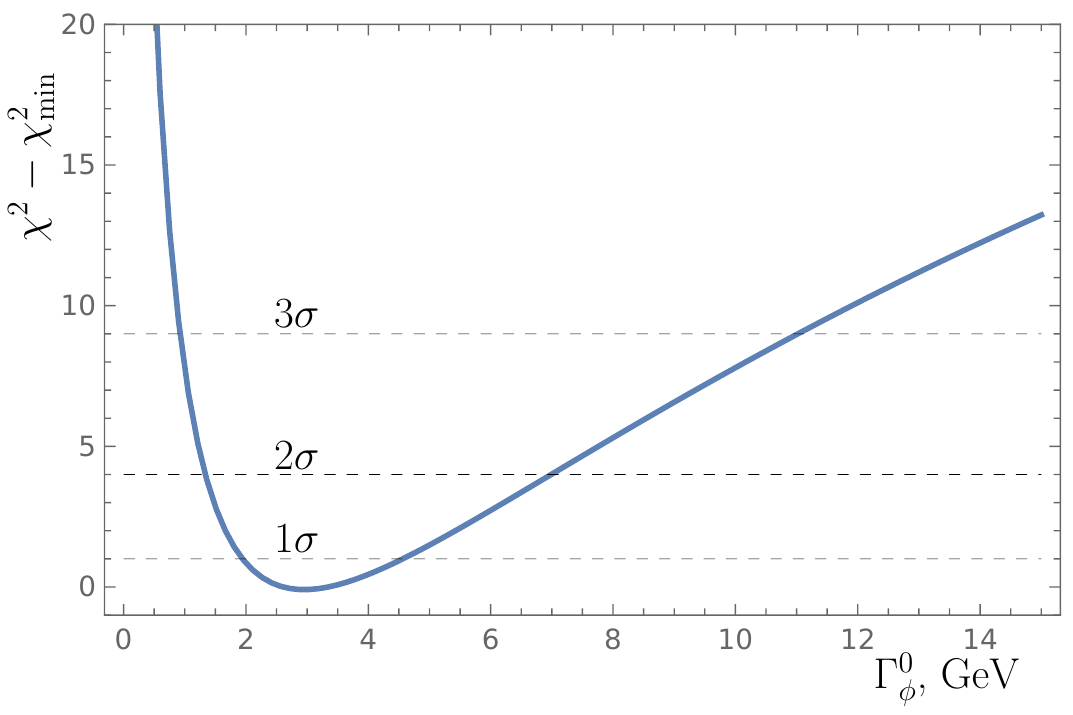}\\
(e) & & (f)\\
\end{tabular}
\caption{The $\chi^2$ distributions.
(a,b,c) --- the two-dimensional (2D) $\chi^2$ distributions
(a) $M_{\phi}$ and $\Gamma^0_\phi$,
(b) $m_{\chi}$ and $\Gamma^0_\phi$,
(c) $M_{\phi}$ and $m_\chi$.
The black dots indicate the ``best values'' of the parameters corresponding to the minimal $\chi^2_{\rm min}=9.02$: $M_\phi=2.4$ GeV, $\Gamma_\phi^0=2.9$ GeV, and $m_\chi=0.42$ GeV.
The value of the third variable is set to its ``best'' value.
(d,e,f) --- the one-dimensional (1D) distributions of $\chi^2$ vs.\
(d): $M_\phi$, (e): $m_\chi$, (f): $\Gamma_\phi^0$. The other two parameters in these plots are set to their best values. }
\label{fig:chi2}
\end{figure}

\begin{figure}[ht]
\centering
\includegraphics[width=9cm]{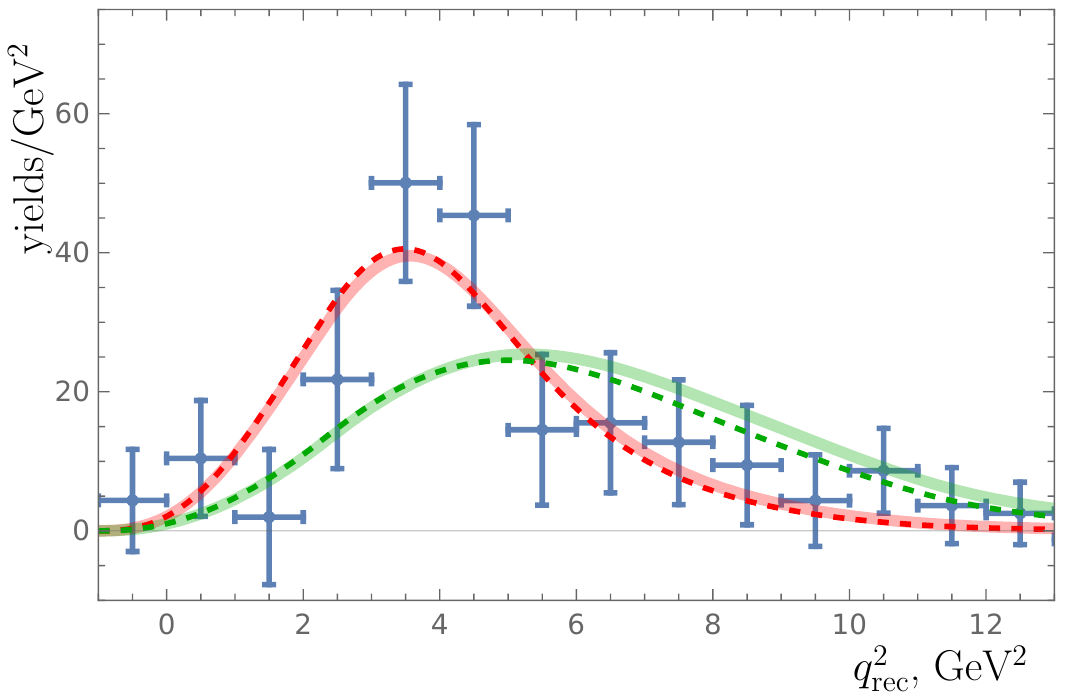}
\caption{\label{fig3}
Belle II data on the decay $B\to K M_X$ (see Fig.~18 from \cite{Belle-II:2023esi} and Fig.~1 from \cite{Fridell:2023ssf}) fitted by our DM model for two different parameter sets:
  the ``best'' point corresponding to the global minimum of $\chi^2$, at
  $M_\phi=2.4$~GeV, $\Gamma^0_\phi=2.9$~GeV  and $m_\chi=0.42$~GeV (red solid curve); 
  a representative point from the $\chi^2$-plateau,
  $M_\phi=20$~GeV, $\Gamma^0_\phi=20$~GeV and $m_\chi=0.42$~GeV (green solid curve).
 The corresponding predictions for $B\to K^* M_X$ are also shown,  assuming the same detection efficiency (red dashed and green dashed curves). Only excess events over the SM expectations are shown.}
\label{fig4}
\includegraphics[width=9cm]{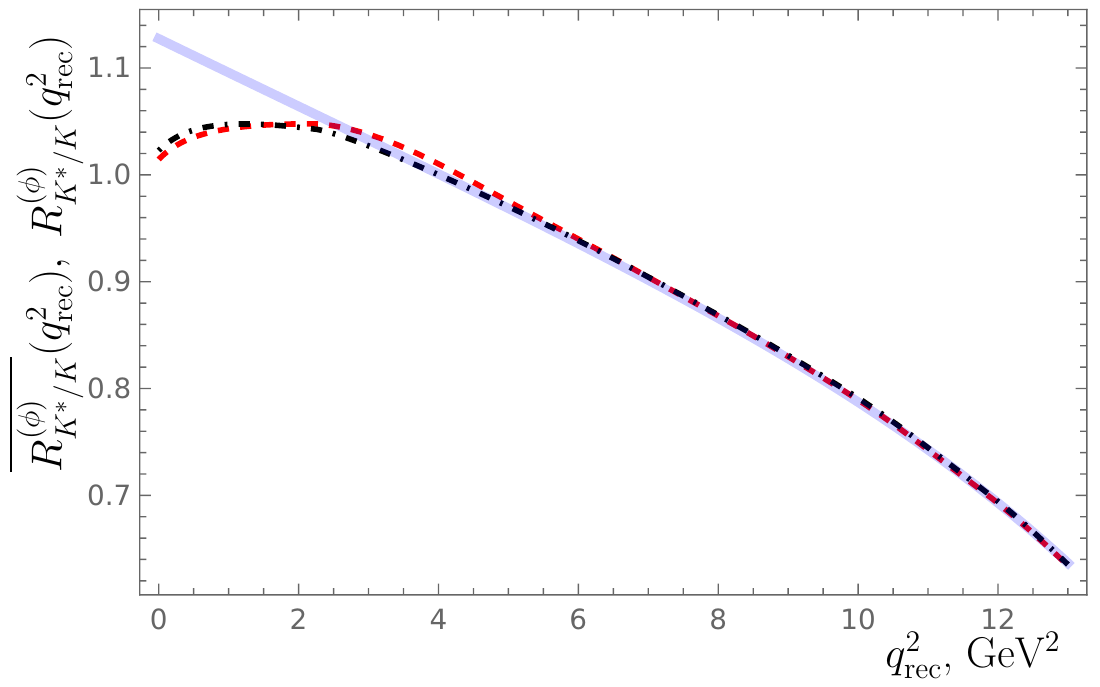}
\caption{The predicted theoretical ratio $R^{(\phi)}_{K^*/K}$ [which is independent of $M_\phi$, $\Gamma^0_\phi$ and $m_\chi$ (blue solid curve)] vs.\ the predicted ``experimentally measurable'' ratio $\overline{R^{(\phi)}_{K^*/K}}$ for different parameter sets [$M_\phi=2.4$~GeV, $\Gamma^0_\phi=2.9$~GeV and $m_\chi=0.42$~GeV (red dashed curve),~and $M_\phi=20$~GeV, $\Gamma^0_\phi=20$~GeV and $m_\chi=0.42$~GeV (black dotted-dashed curve)] vs.\ $q_\mathrm{rec}^2$.}
\label{fig:Kstar_K_ratio}
\end{figure}

The results of fitting the experimental differential distributions  $d\Gamma/dM_X^2$ by our formulas are shown in Fig.~\ref{fig:chi2}, which provides both two- as well as one-dimensional distributions of $\chi^2$.
The 2D plots of Fig.~\ref{fig:chi2}(a,b,c) show the ``best values'' of the DM parameters,
$M_\phi=2.4$ GeV, $\Gamma_\phi^0=2.9$ GeV, and $m_\chi=0.42$ GeV (\emph{the} black dot), corresponding to the global minimum of $\chi^2$ [$\chi^2_{\rm min}=9.02$].
The 2D plots of Fig.~\ref{fig:chi2}(a,b,c) provide also the 1,2,3... $\sigma$ ranges indicating sizeable correlations between the DM parameters. The 1D plots Fig.~\ref{fig:chi2}(d,e,f) show the $\chi^2$ distributions of the individual DM parameters around the ``best'' point leading to the following estimates of the 1$\sigma$ uncertainties:
\begin{eqnarray}
M_\phi=2.4\pm 0.4 \mbox{ GeV},\quad \Gamma^0_\phi=2.9^{+1.1}_{-0.9} \mbox{ GeV}, \quad  m_\chi=0.42^{+0.2}_{-0.4}\mbox{ GeV}.
\end{eqnarray}
The differential distribution $d\Gamma/dM_X^2$ corresponding to the ``best'' parameters is shown as the red line in 
Fig.~\ref{fig3}. Clearly, one obtains a nice description of the shape of the measured spectrum.

The value of $\kappa$ is obtained from the extracted values of $M_\phi$ and $\Gamma_\phi^0$ via Eq.~(\ref{Gamma}):
\begin{equation}
   \label{kappa}
\kappa\approx 5.
\end{equation}
From the total number of $B$ mesons produced at Belle II ($N_\mathrm{tot}=3.99 \cdot 10^8$), and the observed excess of approximately 170 events, and using $\kappa$ from (\ref{kappa}), we can estimate the coupling $g_{b\to s \phi}$ in (\ref{eq:dGdq2}) as
\begin{eqnarray}
g_{b\to s \phi}\sim 7 \cdot 10^{-8}.
\end{eqnarray}

Figure 3 shows our prediction for the differential distributions of the excess events in $B\to K^* M_X$ and $B\to K M_X$ decays. Assuming the same detection efficiency for $B\to K^*M_X$ and $B\to K M_X$ decays, the ratio $\overline{ R^{(\phi)}_{K^*/K}}(q_\mathrm{rec}^2)$ of excess events in the experimentally measured differential distributions in $B\to K^*M_X$ and $B\to K M_X$ decays,
\begin{eqnarray}
\overline{R^{(\phi)}_{K^*/K}}(q_\mathrm{rec}^2)=\frac{\overline{d\Gamma^{\rm eff}}(B\to K^* \chi\bar\chi)/dq_\mathrm{rec}^2}
{\overline{d\Gamma^{\rm eff}}(B\to K \chi\bar\chi)/dq_\mathrm{rec}^2},
\end{eqnarray}
is practically independent of the model parameters, such as $M_\phi$, $m_\chi$, and $\Gamma^0_\phi$, since the dependence on~these parameters approximately cancels out in the ratio. [Recall that the dependence on these parameters cancels exactly in the theoretical ratio $R^{(\phi)}_{K^*/K}$ of Eq.~(\ref{R}).]

Consequently, we have an approximate relation between the measured $\overline{R^{(\phi)}_{K^*/K}}(q_\mathrm{rec}^2)$ and
the theoretical $R^{(\phi)}_{K^*/K}(q_\mathrm{rec}^2)$, which is fulfilled with very high accuracy in a broad range of momentum transfers (see Fig.~\ref{fig:Kstar_K_ratio}):
\begin{eqnarray}
\overline{R^{(\phi)}_{K^*/K}}(q_\mathrm{rec}^2)\simeq R^{(\phi)}_{K^*/K}(q_\mathrm{rec}^2).
\label{eq:Bstar_B_ratio}
\end{eqnarray}
This useful result is not surprising, given the relation in~(\ref{eq:Grec2_Gq2}). Notice, however, that if the reconstruction
efficiencies for the decays $B\to K^* M_X$ and $B\to K M_X$ differ significantly, then (\ref{eq:Bstar_B_ratio}) must be adjusted accordingly.

It is important to note that the experimental $q_\mathrm{rec}^2$ distribution reaches its maximum at $q_\mathrm{rec}^2 \approx 5 \mbox{ GeV}^2$, where $R^{(\phi)}_{K^*/K}(q_\mathrm{rec}^2)\approx 1$. This observation implies that the number of excess events in the decays $B\to K^* M_X$ and $B\to K M_X$ --- assuming the same detection efficiency for both processes --- should be approximately equal in magnitude and the corresponding differential distributions in $q_\mathrm{rec}^2$ should have similar shapes. This behavior is illustrated in Fig.~\ref{fig3}, where predictions~for the expected excess events for the $B\to K^* M_X$ decay are represented by dashed curves.

Before closing this section, let us make the following remark:
In the 2D plots of Fig.~\ref{fig:chi2}, one can identify also a region of large $M_\phi$ corresponding to a plateau
  in $\chi^2$ which provides a formally still acceptable description of the data at the 3$\sigma$ level.
  In this region, the $\phi$ propagator becomes practically insensitive to $q^2$ (from the $B$-meson kinematical
  decay region) and reduces to a constant, such that the shape of the spectrum practically loses its sensitivity
  to $\Gamma_\phi$ and $m_\chi$.  The green line in Fig.~\ref{fig3} presents $d\Gamma/dM_X^2$ for
  $M_\phi=20$ GeV, $\Gamma_\phi=20$ GeV,  $m_\chi=0.42$ GeV.
  (In practice, for $M_\phi>15$ GeV any value of $\Gamma_\phi$ and $m_\chi$ may be used for drawing this plot.)
  This region of the parameter space of the DM does not seem interesting from the physical point of view; also the
  shape of the spectrum does not fit the data well. Nevertheless, since the region is still compatible with the data
  with 3$\sigma$ accuracy, we mention this region as a ``marginal'' region.

\section{Conclusions}
In the current study, our aim was limited to the analysis of the impact of the enhancement in the $B\to K M_X$ decay, observed by Belle II~\cite{Belle-II:2023esi}, on some DM scenario involving a scalar mediator. For this model, the following conclusions, based solely on Belle II data, hold:
\begin{itemize}
\item
  The model discussed is consistent with the experimental data and provides a good description of the shape
  of the excess observed. One can clearly identify the set of the ``best'' values of the DM parameters
  \begin{eqnarray}
  M_\phi=2.4\pm 0.4~{\rm GeV}, \quad \Gamma^0_\phi=2.9^{+1.1}_{-0.9}~{\rm GeV}, \quad m_\chi=0.42^{+0.2}_{-0.4}~{\rm GeV}.
  \nonumber
  \end{eqnarray}
\item
Both the shape and the normalization of the ratio of the excess events which may be measured experimentally,
$\overline{ R^{(\phi)}_{K^*/K}}(q_\mathrm{rec}^2)$, may be well approximated by the theoretical ratio (\ref{R}) evaluated
at the same $q_\mathrm{rec}^2$, $R^{(\phi)}_{K^*/K}(q_\mathrm{rec}^2)$:
\begin{eqnarray}
\nonumber
\overline{R^{(\phi)}_{K^*/K}}(q_\mathrm{rec}^2)\simeq R^{(\phi)}_{K^*/K}(q_\mathrm{rec}^2).
\end{eqnarray}
Consequently, the shape of $\overline{ R^{(\phi)}_{K^*/K}}(q_\mathrm{rec}^2)$ is largely independent of the DM model parameters --- such as $M_\phi$, $\Gamma_\phi$ and $m_\chi$ --- and provides a clear signature of the DM scenario based on the scalar-mediator field.
\item
Within the framework of the model considered, the yields of the excess events over the SM background for the decays $B\to K^* M_X$ and $B\to K M_X$ should be approximately equal in magnitude and exhibit a qualitatively similar shape in the $q_\mathrm{rec}^2$ distribution, assuming the same detection efficiency for both reactions.
\end{itemize}
These constraints on DM particles may be combined with constraints coming from other phenomena (see, e.g., \cite{Abdughani:2023dlr})~but such kind of analysis is beyond the scope of our interest in this analysis.

\vspace{.2cm}\noindent
{\it Acknowledgments.}
The research was carried out within the framework of the program \emph{Particle Physics and Cosmology}
of the National Center for Physics and Mathematics. The authors thank K\aa{}re Fridell for helpful
comments on the study~\cite{Fridell:2023ssf}.

%
\end{document}